\documentstyle{article}
\renewcommand{\baselinestretch}{2}
\begin{document}
\begin{sloppypar}
\large
\begin{center}
{\bf THE TIGHT-BINDING APPROACH TO THE DIELECTRIC RESPONSE IN THE
MULTIBAND SYSTEMS }

\vspace{5mm}

P. \v{Z}upanovi\'{c}

{\em Department of Physics,
Faculty of Science and Art, University of Split,
Teslina 10, 58000 Split, Croatia} \vspace{5mm}\\
 A. Bjeli\v{s} and  S. Bari\v{s}i\'{c}

{\em Department of Physics,
Faculty of Science, University of Zagreb,
P.O.B. 162,

41001 Zagreb, Croatia}
\end{center}
\vspace{5mm}

\begin{center}
{\bf Abstract}
\end{center}

\mbox{}

Starting from the random phase approximation for the weakly coupled multiband
 tightly-bounded electron systems,
we calculate  the dielectric
matrix in terms of intraband and interband transitions.
The advantages of this  representation
 with respect to the usual plane-wave  decomposition are
pointed out.  The analysis becomes  particularly transparent in the
long wavelength limit, after performing the multipole expansion of bare Coulomb
matrix elements. For illustration, the collective modes and the macroscopic
dielectric function for a general cubic lattice are derived. It is shown
that the dielectric instability in conducting narrow band systems proceeds by
a common softening of one transverse and one longitudinal mode.
Furthermore, the self-polarization corrections which appear in the
macroscopic dielectric function for
finite band systems, are identified as a combined effect of intra-atomic
exchange interactions between electrons sitting in different orbitals and
a finite inter-atomic tunneling.

\bigskip


\bigskip


\newpage

\section{ Introduction}
\mbox{}

Random phase approximation (RPA) is the well-known textbook method \cite{pines}
for treating the dielectric properties of the
weakly interacting electron liquids. Unfortunately the
widely used [2 - 5] extension
of this method to the multiband systems leads to a
cumbersome  problem, even when limited only to the
calculation of collective modes of the tightly bound electrons
in the optical regime. In the present text we propose therefore
a new formulation of RPA which leads to a more
convenient description of the dielectric response of the tightly bound
electrons. In particular it  greatly simplifies the analysis of the
collective modes giving a direct insight into their physical
origin. It also resolves some long lasting controversies
\cite{adler,onodera,sinha}
on the local field effects in multiband insulators and metals.

The paper is organized as follows. In Sec.2 we formulate the
problem and derive the
tight-binding expression for the dielectric matrix. In Sec.3
the determinant of this matrix for the
lattice with the cubic symmetry is explicitly calculated in the
long wavelength limit in which one can make the multipole expansion of
the long range contributions to the bare Coulomb matrix elements. The
collective modes and the
macroscopic dielectric function which follow from this result are
considered in Secs. 4 and 5 respectively. The  latter Section also
contains the discussion of the so-called self polarization
corrections. Concluding remarks are given in Sec.6.

\newpage

\section { Dielectric matrix in the tight binding approach}
\mbox{}

    In the standard approach \cite{adler,wiser} one uses the
plane wave representation and starts from the
infinite dielectric matrix
 $\varepsilon_{\mbox{\bf \small G,G \large}'}(\mbox{\bf q}, \omega)$
 with {\bf G}
and {\bf G}' denoting all the vectors of the reciprocal lattice. In order to
determine the collective modes and the macroscopic dielectric
function
 $\varepsilon_{M}^{-1} = ( \varepsilon^{-1})_{\mbox{\bf \small 00 \large}}$
one has to calculate the
determinant of the dielectric matrix $\varepsilon_{\mbox{\bf \small G,G
\large}'}$.
Unfortunately there was no straightforward way
to reduce the dimension of this determinant even
when only a finite number of presumably relevant tightly bound
bands was taken into account. The reason is
that the plane wave basis used for the matrix elements of the
Coulomb interaction is not suitable for the description of the
tightly bound electronic band states. As a consequence, already in the
extreme long wavelength limit the analysis of the dielectric properties was
rather complicated \cite{adler},
while sophisticated and physically  nontransparent numerical
algorithms were inevitable for more complex situations \cite{lcc,hl}.

  One easily appreciates that the tight binding (TB) Wannier-like
representation with explicit band indices is more appropriate than
the plane wave one.  Having in  mind the fact that the basis
defined by band indices and the values of wave vector within the
first Brillouin zone brings in
 the full  crystal symmetry (including
the discrete translations)  like the basis defined in the extended
zone scheme,  we introduce
in the present work the dielectric matrix in the former basis, without
any reference to the plane wave indices  {\bf G}.  This approach was
already used in studying the dielectric  properties of the single metallic
band \cite{barisic}, but not, to the best of our knowledge, for multiband
systems. In fact,  the TB basis was frequently used indirectly
\cite {adler,wiser,onodera,sinha,hanke}, {\em i. e.} in the calculation
of the RPA multiband polarizabilities which appear
in the matrix elements $\varepsilon_{\mbox{\bf \small G,G \large}'}$.
This led to some simplifications in inverting the dielectric  matrix
$\varepsilon_{\mbox{\bf \small G,G \large}'}(\mbox{\bf q}, \omega)$
\cite {hanker}. We point out below some crucial advantages of  the present
direct TB representation.

We start by calculating the mean values of the components
\begin{equation}
\rho_{ll'}(\mbox{\bf q},t) \equiv 2\sum_{\mbox{\bf k}}<a^{+}_{l}(\mbox{\bf
k},t)
a_{l^{'}}(\mbox{\bf k+q},t)>
\label{1}\end{equation}
of the density response
to the corresponding components of the external potential
$V^{ext}(\mbox{\bf r},t)$,
\begin{equation}
V_{ll'}^{ext}(\mbox{\bf q}, t)
=\frac{1}{N} \sum_{\bf R}e^{i \mbox{\bf q}   \mbox{\bf R}}
 \int d^{3} r \; \varphi^{*}_{l}(\mbox{\bf r-R}) V^{ext}(\mbox{\bf r},t)
 \varphi_{l^{'}}(\mbox{\bf r-R}).
\label{2}
 \end{equation}
Here $\varphi_{l}(\mbox{\bf r-R})$ is the TB orbital for the $l$-th
band at the {\bf R}-th site, and $a_{l}^{+}(\mbox{\bf k})$ is the
creation operator for the corresponding Bloch state.
The present discussion is limited to the weak coupling
 in which the
bare intraband and interband Coulomb interactions are smaller than the
bandwidths and  the interband energy differences respectively.
Starting from the equations of motion for the operators of the
electron-hole pairs $a_{l}^{+}(\mbox{\bf k}) a_{l^{'}}(\mbox{\bf k+q})$,
and performing the standard RPA steps \cite{pines,nozieres},
 one arrives at the system of linear equations
\begin{equation}
\sum_{r}(\delta_{pr}-V_{rp} \Pi_{p})\rho_{r}=\Pi_{p}V_{p}^{ext}.
\label{3}
\end{equation}
The indices in the eq.(3) stay for the ordered pairs of band indices,
$p=(l,l')$
 and denote the
 transitions between the tight-binding bands.  Note that the spin
indices can be omitted in the RPA. The
function $\Pi_{p}(\mbox{\bf q}, \omega)$ is the RPA
polarization diagram
\begin{equation}
\Pi_{p}({\bf q},\omega)=\frac{2}{N}
\sum_{\mbox{\bf k}} \frac{n_{l^{'}}(\mbox{\bf k})-n_{l}(\mbox{\bf k+q})}
{\omega-E_{l}({\bf k}+{\bf q})+E_{l^{'}}({\bf k})+i\eta}
\label{4}
\end{equation}
 with the occupation and dispersion of the $l$-th band given by $n_{l}$
and $E_{l}$ respectively. Evidently, $\Pi_{p}$ is finite only for transitions
between (partially) full and (partially) empty bands
as well as for the transitions within the partially filled bands.

 $V_{pr}(\mbox{\bf q})$ in eq.(\ref{3})
is the matrix element of the bare Coulomb interaction
in which we keep only the contributions with two and two
TB orbitals centered on same crystal sites (two center integrals),
  \begin{equation}
 V_{pr}({\bf q})=\sum_{\mbox{\bf R} }
e^{i{\bf qR}} \int d^{3} r \int d^{3} r'
\varphi^{*}_{l_{1}}({\bf r}-{\bf R})\varphi^{*}_{l_{2}}({\bf r'})
\frac{e^{2}}{\mid{\bf r}-{\bf r'}\mid}
\varphi_{l_{1}^{'}}({\bf r}-{\bf R})
\varphi_{l_{2}^{'}}({\bf r'})
\label{5}                                                           ,
\end{equation}
with  $ p=(l_{1}, l_{1}')$ and $ r = (l_{2}, l_{2}')$.
This usual TB approximation which greatly simplifies the system of
equations (\ref{3}) and works best for long wavelengths \cite{barisic},
is based on the assumption that the overlaps  between
the orbitals on the neighboring sites are so small that the
corresponding contributions to the Coulomb matrix elements
are negligible with respect to the contributions retained in the eq.(\ref{5}).
Basically, this is the only restriction on the present RPA approach.
In return it leads to simple $p, r$ selection rules, as will be
discussed below. Further on, in the long wavelength limit it will be
sufficient to keep only the leading
contributions in the multipole expansion of the two-site
$(\mbox{\bf R} \neq 0)$ terms in eq.(\ref{5}).
The on-site $(\mbox{\bf R} = 0)$ terms are usually omitted, or at least not
treated explicitly in the weak coupling limit. However, as it will be
argued below, some of these terms are relevant for the understanding
of the local field effects.

In the present approach the dielectric matrix is given by
the coefficients on the left-hand side of eq.(\ref{3}),
\begin{equation}
\varepsilon_{p,r}(\mbox{\bf q},\omega)= \delta_{p,r}-V_{rp}\Pi_{p},
\label{6}
\end{equation}
in accordance with the definition which follows from the Dyson equation
for the screened Coulomb interaction \cite{zbb}, and with the standard
definition for the special single band case \cite{pines}.

With the wave vector {\bf q} restricted to the first
Brillouin zone and with $l$ covering all band indices, the
matrix $\varepsilon_{p,r}(\mbox{\bf q},\omega)$ represents, like the matrix
$\varepsilon_{\mbox{\bf \small G,G \large}'}(\mbox{\bf q},\omega)$,
a complete RPA dielectric response,
with all local field effects taken into account. Strictly, its dimension
is infinite, due to the infinite number of possible interband
transitions with $\Pi_{p} \neq 0$. However, an approximation
which  neglects interband transitions with small polarizations
$\Pi_{p}$ [{\em e.g.} those from the deep (core) orbitals and those to very
high empty bands], reduces the dimension of
$\varepsilon_{p,r}(\mbox{\bf q},\omega)$ to a
finite value, equal to the number of retained interband and intraband
transitions.  This type of simplification  with the explicit
physical justification cannot be achieved in the plane
wave representation which requires numerical truncations in   both
band and reciprocal lattice indices \cite{lcc,hl,van}.

\section { Long wavelength dielectric response of the cubic lattice}
\mbox{}

   The further advantage of the dielectric matrix in the  TB representation
becomes apparent in the explicit long wavelength limit. For the sake of
definitiveness let us consider the cubic lattice and at first take into account
only the two-site terms in the eq.(\ref{5}). An analogous procedure can
be carried out for any crystal symmetry. We distinguish the following
three types of dominant lattice sums \cite{cohen}
in the multipole expansion of the various matrix elements
$V_{pr}(\mbox{\bf q} \rightarrow 0)$.

 For both $p$ and $r$ representing intraband
transitions [{\em i.e.} the transitions within partially filled bands, with
$p=(l_{1}, l_{1})$ and $r=(l_{2},l_{2})$], the
dominant contribution is the monopole-monopole one, given by \cite{tekst1}
\begin{equation}
V_{pr}({\bf q})= \frac{4\pi e^{2}}{a^{3}q^{2}},
\label{7}
\end{equation}
where $a$ is the lattice constant.

 If both indices represent interband
transitions [{\em i.e.} $p=(l_{1}, l_{1}'$), $r= (l_{2}, l_{2}')$ with
 $l_{1} \neq l_{1}' $
and $l_{2} \neq l_{2}' $] the lowest possible multipole contribution is
the dipole-dipole term
\begin{equation}
V_{pr}=\frac{8\pi}{3}\frac{\mu_{p\parallel} \mu_{r\parallel}}{a^{3}}-
\frac{4\pi}{3}
\frac{\mbox{\boldmath $\mu$}_{p\perp} \mbox{\boldmath $\mu$}_{r\perp}}{a^{3}}.
\label{8}
\end{equation}
Here
\begin{equation}
\mbox{\boldmath $\mu$}_{p}=e\int d\mbox{r}^{3}\,
\varphi_{l}^{*}(\mbox{\bf r})\mbox{\bf r}\varphi_{l^{'}}
(\mbox{\bf r})
\label{9}
\end{equation}
is the dipole matrix element for the transition from the $l$-th
to the $l'$-th local function, and $\mu_{p\parallel}$
and $\mbox{\boldmath $\mu$}_{p\perp}$ are its
respective projections parallelly and perpendicularly to the
wave vector.

The third type of matrix elements are those with one interband
and one intraband  transition [{\em i.e.}  $p=(l_{1},l_{1}')$
 with $l_{1}, \neq l_{1}'$ and
   $r=(l_{2}, l_{2})$ , or  vice versa] for which
the lowest
multipole contribution is of the monopole-dipole form,
\begin{equation}
V_{pr}=-\frac{4\pi ie}{a^{3}}\frac{\mu_{p \parallel}}{q} .
\label{10}
\end{equation}

We emphasize that the long-range part of the Coulomb interaction contributes
only to the above three types of matrix elements
 in the long wavelength limit $\mbox{\bf q} \rightarrow 0$.
The other terms in the
multipole expansions are higher orders in components of {\bf q}, and as such
are vanishing  in this limit. They are therefore irrelevant even if {\em e. g.}
the dipolar factors
in the eqs.(\ref{8}) and (\ref{10}) are absent for symmetry reasons.

   Let us now divide all transitions into four sets, taking into  account
particular properties of the cubic symmetry. Each dipolar matrix element
(\ref{9}) is directed along one of three fourfold
 rotation axes (x, y and z)
\cite {ll}. All  interband (dipolar) transitions can be  therefore divided
into three sets, $\{p_{j}\}$, where $j = 1,2, \mbox{ and }3 $
 stay for $ x,y$ and $z$ axis respectively. The fourth set $\{p_{0}\}$
contains all (at most three \cite{tekst1})
 intraband (monopolar) transitions. Furthermore, for each dipolar matrix
element  $\mbox{\boldmath $\mu$}_{p_{1}}$
 there are two other matrix elements
 $\mbox{\boldmath $\mu$}_{p_{2}}$ and $\mbox{\boldmath $\mu$}_{p_{3}}$
 with the same absolute
value, which complete a three-dimensional irreducible
representation of the cubic point group \cite{ll}.

These symmetry properties, together with
the factorized form of the matrix elements
(\ref{8}) and (\ref{10}), reduce the system of eqs.(\ref{3}) to four
linear equations. To this end we introduce the following linear
combinations of the density
components $\rho_{p}$ belonging to the four sets defined above:
\newcommand{\be}{\begin{equation}}
\newcommand{\ee}{\end{equation}}
\be
\label{12a}
\rho_{0}=\sum_{p \in \{p_{0}\}} \rho_{p}
\ee
and
\be
\label{12b}
\rho_{j}=-i
\frac{\mbox{\bf q}  \hat{\mbox{\bf n}}_{j} }{e}
\sum_{p \in \{p_{j}\} }
\mu_{p} \rho_{p} \hspace{20mm}  j=1,2,3
\ee
with $\hat{\mbox{\bf n}}_{j} $ denoting the unit vector along the $j-th$
 axis.
Summing up separately equations (\ref{3}) with $ p \in \{p_{0}\}$ and
$p \in \{p_{j}\} , j = 1, 2,3$, one
arrives after few straightforward steps to the decoupled systems of two
and two equations. The first system
represents the longitudinal response

\be
\label{13}
[1+(4\pi \alpha_{c})^{-1}] \rho_{0} +\sum_{j=1}^{3} \rho_{j}=-
\frac{a^{3}q^{2}}{4\pi e^{2}} V^{ext},
\ee
\be
\label{15}
\rho_{0}+\left[ \frac{2}{3} +(4 \pi \alpha_{I})^{-1} \right]
\sum_{j=1}^{3} \rho_{j}=
i\frac{a^{3}}{4\pi e}\mbox{\bf q} \cdot \mbox{\bf E}_{ext},
\ee
while two linearly independent equations among the following
three equations

\be
\left[(4 \pi \alpha_{I})^{-1} -\frac{1}{3}\right]
\left[ \frac{\hat{\mbox{\bf q}}  \hat{\mbox{\bf n}}_{j}}
{\hat{\mbox{\bf q}}  \hat{\mbox{\bf n}}_{i}}\rho_{i}-
\frac{\hat{\mbox{\bf q}}  \hat{\mbox{\bf n}}_{i}}
{\hat{\mbox{\bf q}}  \hat{\mbox{\bf n}}_{j}}\rho_{j} \right]=
\frac{ia^{3}}{4\pi e} ( \mbox{\bf q} \times \mbox{\bf E}_{ext}) \cdot
 \hat{\mbox{\bf n}}_{k}
\label{16}
\ee
(with {$i \neq j \neq k$) represent the transverse response.
Here $\hat{\mbox{\bf q}} \equiv \mbox{\bf q}/q$,
$V^{ext}(\mbox{\bf q}, \omega)$ is the external scalar potential and
$\mbox{\bf E}_{ext}(\mbox{\bf q}, \omega)$ is the external
electric field. Eqs.(\ref{16}) are slightly generalized with respect to the
original eqs.(3), since they  allow for the finite transverse electrical
field. The intraband  and interband  polarizabilities are given by
\be
\label{14a}
 \alpha_{c}=-\frac{ e^{2}}{a^{3}q^{2}}\sum_{p \in \{p_{0}\} }\Pi_{p}
\ee
\be
 \alpha_{I}=-\frac{1}{a^{3}}\sum_{p \in \{I\}}   |\mu_{p}|^{2} \Pi_{p}
\label{14b}
\ee
respectively. $\{I\}$  stands for any of three sets $\{p_{j}\}, j=1,2,3.$

The above separation to the longitudinal and transverse response
does not depend on
the direction of {\bf q}, as it should be for the cubic crystal.
Since the original density components $\rho_{p}$ follow from
eqs.(\ref{3}) once
the combinations which figure in eqs.(13 - 15)
 are known, the above procedure
gives the explicit solution for the dielectric response of cubic crystals.
In particular,  since the systems of equations (\ref{3}) and (13 - 15)
are connected by linear
transformations and thus have a common determinant, it follows immediately from
the latter
equations that
\be
\label{17}
det [\varepsilon_{p,r}(\mbox{\bf q},\omega)] =
\varepsilon_{l}(\varepsilon_{t})^{2}
\ee
with
\be
\label{18a}
\varepsilon_{l} = (1+4\pi \alpha_{c}) \left( 1+\frac{8\pi \alpha_{I}}{3}
\right)
 - 16 \pi^{2}\alpha_{c}\alpha_{I}
\ee
and
\be
\label{18b}
\varepsilon_{t} = 1- \frac{4\pi \alpha_{I}}{3}.
\ee
The original determinant
of infinite order is reduced so to the algebraic expression in which the
infinite summation in $\alpha_{I}$ can be easily truncated due to the
explicit physical meaning of the terms in eq.(\ref{14b}).

\section { Collective modes}
\mbox{}

The expressions (18 - 20) are particularly convenient
for the  discussion
of  the collective modes which appear as isolated poles of $\det(\varepsilon)$
in
the $\omega$-plane, and
of their Landau damping due to the incoherent intraband and interband
electron-hole excitations, provided the band dispersions and corresponding TB
orbitals are specified. The details of this calculation for simple band
models are discussed elsewhere \cite{zbb}. Here we
present the most important conclusions concerning the collective modes.

Besides the obvious case of a single band conductor
$\alpha_{C} \neq 0,   \alpha_{I} = 0$ with intraband plasmons as collective
excitations,
the result
(\ref{17}) covers also a less evident asymptotic limit of an atomic (molecular)
insulator. Namely, after assuming that all interband energies
$|E_{l}^{'} - E_{l}| \equiv E_{p}$
are much larger than the corresponding bandwidths, the
interband polarizability (\ref{14b}) reads
\be
\label{19}
 \alpha_{I} =-\frac{2}{a^{3}}\sum_{p \in \{I\}}
\frac{ n_{p}E_{p}|\mu_{p}|^{2}}{\omega^{2}-E_{p}^{2}}
\ee
{\tt where $n_{p}$ is the number of electrons per a site that take part in
$p$-th transition.}
Inserting this expression  together with $\alpha_{c} = 0$ into
eqs.(\ref{18a},\ref{18b}),
one gets the dipolar collective modes whose spectrum just coincides with
that of Frenkel excitons \cite{knox}.  It should be noted in this respect that
the present RPA result
is valid in the weak coupling regime \cite{we}. We remind that the same
spectrum is obtained \cite{anderson} in the opposite strong coupling regime
when the bandwidths are negligible  with respect to the
on-site Coulomb repulsion $V_{pr}(\mbox{\bf R}=0)$ between two electrons
on different orbitals ({\em i. e.} the
$\mbox{\bf R}=0$ contribution in the expression (\ref{5})
with $l_{1}=l_{1}^{'}$ and $l_{2}=l_{2}^{'}$).

The Frenkel excitonic spectrum thus appears to be a common
asymptotic limit of the two otherwise incompatible regimes.
Indeed, the corrections due to the finite bandwidths
are scaled differently in the two regimes, i.e. by $E_{p}$ and
$V_{pr}(\mbox{\bf R}=0)$ respectively.
In the strong coupling regime the collective modes
are coherent superpositions of atomic
electron-hole excitations, gradually delocalized by finite
bandwidths. The appropriate representation in the weak
coupling regime starts from the electrons in Bloch states. Accordingly,
in contrast to the former, the latter regime includes also the
narrow band conductors in which the interband polarizability can
be still approximated by eq.(\ref{19}), while the intraband polarizability
is finite and given by
\be
\label{20}
4\pi \alpha_{c} =    -\frac{\omega_{pl}^{2}}{\omega^{2}}           ,
\ee
where $\omega_{pl} =4\pi n_{e} e^{2}/m^{*}a^{3}$
 is the frequency of the intraband plasmon, and $m^{*}$ and $n_{e}$
 are respectively the effective mass and the number of electrons
(or holes) in the  metallic band(s).

As can be seen from eq.(\ref{18a}), the finiteness of $\omega_{pl}$
causes the renormalization of the dipolar longitudinal
modes, but does not affect the transverse modes (\ref{18b}).
This coupling between longitudinal intraband and interband collective modes
is to be traced back to the finite monopole-dipole interaction (\ref{10}).
Although the transverse modes (\ref{18b}) are not screened by the intraband
charge fluctuations, there is an interesting relation between
them and the hybridized longitudinal modes of eq.(\ref{18a}).
Let the frequency
of the lowest transverse mode in eq.(\ref{18b}),
$\omega_{t_{low}}$, tend to zero. Then
the equation $\varepsilon_{l} =0$ has the solution, $\omega_{l_{low}}$,
which also tends to zero.
The reverse is also true. The ratio of two frequencies is
\be
\label{21}
\left(\frac{\omega_{l_{low}}}{\omega_{t_{low}}} \right)^{2}=
\frac{\omega_{pl}^{2}}{\omega_{pl}^{2}+
 \frac{ 4 \pi \alpha_{I}(\omega=0) \prod_{p}E_{p}}{\prod_{j}
\omega_{T_{j}}^{2}}}
\ee
where $\omega_{T_{j}}$ are frequencies of all  other (stable)
transverse collective modes.
Note that $\omega_{l_{low}} < \omega_{t_{low}}$. The dielectric
instabilities in multiband conductors are
thus characterized by the simultaneous
softening  of two collective modes. This somewhat surprising,
and to the best of our knowledge new result is the consequence
of the proper treatment of the  metallic screening
of the interband (dipolar) excitations in the present approach.
We note that the cubic symmetry chosen here is not essential for the
validity  of this result \cite{zbb}.}

\section { Macroscopic dielectric function}
\mbox{}

The macroscopic dielectric function $\varepsilon_{M}$ follows immediately
from eqs.(\ref{13},\ref{15}). To this end it suffices to utilize the averaging
scheme \cite{wiser} by which the probe electrical field and the corresponding
density response are
$\mbox{\bf E}_{ext} =i\mbox{\bf q}V^{ext}(\mbox{\bf q})/e$
and $\rho(\mbox{\bf q})= \sum_{j=0}^{3}\rho_{j}(\mbox{\bf q})$ respectively.
According to the standard definition \cite{pines}
$\varepsilon_{M}$ is then given by
\be
\label{24}
\varepsilon_{M}(\mbox{\bf q},\omega) =
 [1+\frac{\rho(\mbox{\bf q},\omega)}
{a^{3} q^{2} V^{ext}/4 \pi e^{2}}]^{-1} =
\frac{\varepsilon_{l}}{\varepsilon_{t}}=
 1 + 4\pi \alpha_{c} +\frac{4 \pi \alpha_{I}}
{1-\frac{4 \pi}{3}\alpha_{I}}.
\ee
This expression reproduces some earlier results
\cite{adler,wiser,onodera,sinha}, but, unlike them,
does not contain any type of the so-called self-polarization corrections.
In our approach these corrections are related to  the on-site (intra-atomic)
contributions to the bare Coulomb matrix elements (\ref{5}), not taken into
account
until now.
Obviously, all these contributions are independent of {\bf q}.  Furthermore,
it is
easy to see that due to the symmetry reasons only those on-site matrix
elements  $V_{pr}(\mbox{\bf R}=0)$,  with
 $p$ and $r$ representing the transitions of the same (monopolar,  dipolar,
etc)
type,  are finite. On the other side,
in the  RPA it is sufficient to keep only those on-site terms for which
the leading multipole contributions in the corresponding two-site sums are
constant or vanish.
Thus, the on-site terms which complete the monopole-monopole two-site  sums
(\ref{7}) in eq(\ref{5}) [like {\em e.g.} the  on-site "electrostatic"
repulsions
$V_{pr}(\mbox{\bf R}=0)$ between two electrons sitting at the same or different
orbitals] are irrelevant in the limit $\mbox{\bf q} \rightarrow 0$.
Since the on-site contributions which correspond to the monopole-dipole
two-site sums
(\ref{10}) vanish,  the most
interesting on-site contributions are those going together with the next in
order
( {\em i.e.}  dipole-dipole) two-site contributions (\ref{8}).

The inclusion of those on-site contributions into the
expression (\ref{2})  prevents unfortunately the reduction of the system
(\ref{3}) to the explicitly solvable form (\ref{13},\ref{15}). Still, some
interesting conclusions can be drawn for
simple models \cite{zbb}, like that with only two bands ($l = l_{1}, l_{2}$)
connected by a finite dipole matrix element  (\ref{9})
$\mbox{\boldmath $\mu$}_{p}, p = (l_{1},l_{2})$.
Then the dipole-dipole matrix element $V_{p,\tilde{p}}$ of eq(\ref{8})
has to be completed by the on-site term

\be
\label{25}
 V_{p,\tilde{p}}(R=0) = \int d^{3} r \int d^{3} r'
\varphi^{*}_{l_{1}}({\bf r})\varphi^{*}_{l_{2}}({\bf r'})
\frac{e^{2}}{\mid{\bf r}-{\bf r'}\mid}
\varphi_{l_{2}}({\bf r})
\varphi_{l_{1}}({\bf r'})
\ee
which has the meaning of the exchange between two atomic orbitals.
Furthermore, the interband polarizability (\ref{14b}) reduces to
\be
\alpha_{I} = -\frac{1}{a^{3}}|\mu|^{2}(\Pi_{p}+\Pi_{\tilde{p}}).
\label{25a}
\ee
The macroscopic dielectric
function $\varepsilon_{M}$ can be still written in the form (\ref{24}),
but with the interband polarizability $\alpha_{I}$ replaced by
\be
\label{26}
\alpha_{I}^{'} =
\frac{\alpha_{I}}{1 - V_{p,\tilde{p}}(R=0)(\Pi_{p}+\Pi_{\tilde{p}})} \; \;.
\ee
Note that there is no effect of this type on the intraband (metallic)
polarizability
$\alpha_{c}$, in accordance with the already mentioned fact that
for small {\bf q} the long-range monopole-monopole Coulomb interaction
dominates over all on-site "electrostatic" repulsions.

The denominator in the eq.(\ref{26}) can be simply interpreted
as the atomic screening of the interband
polarizability $\alpha_{I}$. Since it has itself the RPA form,
the formulation (\ref{3})  apparently treats both, the
band and the local (intra-atomic) dielectric responses selfconsistently  at
the same (RPA) level of approximation.

When the bandwidths are finite, the intra-atomic and the inter-band
screenings in  eq.(\ref{26}) cannot be decoupled.
As a consequence the result for   $\varepsilon_{M}$  deviates from the
Lorentz-Lorenz (LL)
form in the sense that the  effective interband polarizability
$\alpha_{I}^{'}$  which enters into eq (\ref{24})
cannot be reduced to the form  (\ref{14b}),  {\em i. e.}  to the
sum of interband polarization diagrams. This  deviation  is usually
named a self-polarization correction,  originally derived by Adler \cite
{adler}
within the plane-wave representation.
The result (\ref{26}) brings us to the clear microscopic interpretation
of this correction, recognized as a combined effect of the
intra-atomic screening due to the on-site exchange Coulomb
interactions and of the interatomic tunneling ({\em i. e.}  of
the finite bandwidths).
Note furthermore that some later results for the macroscopic dielectric
function \cite {wiser,onodera,sinha} cannot be represented in the form
(\ref{26}). The reason might be in the additional approximations of these
works which, in contrast to the present approach, do not
include fully the local RPA on-site screening into the  dielectric matrix.

In the asymptotic limit of
zero bandwidths  the expression for  $\varepsilon_{M}$
reduces  again  to the LL form,  but with the
initial ({\em e. g.} Hartree-Fock) values of the
dipole matrix element ${\mu}$ and the energy difference $E_{p}$
replaced by $|\mu_{eff}|^{2}=|\mu|^{2}E_{p}/E_{p, eff}$
and $E^{2}_{p, eff}=E_{p}^{2} + 2n_{e}E_{p}V_{p,\tilde{p}}(R=0)$
respectively.
In the present approach this renormalization is performed within the RPA
scheme,
{\em i. e.} the result (\ref{26}) includes    both  the
crystal and the on-site dielectric screening on the same RPA level.  It is
appropriate to remind here that with vanishing bandwidths such approach is
valid only if the interactions are small with respect to $E_{p}$.
It can be also noted that Adler \cite {adler}  uses the term
"self-polarization correction" for the on-site contributions to the dielectric
screening even in the zero-bandwidth limit.

The inclusion of the intra-atomic Coulomb
interactions does not alter the previous result (\ref{21}) concerning the
common
instability of the transverse and longitudinal modes in narrow band conductors.
{\em E. g}, in the asymptotic limit of zero bandwidths we get, after
the inclusion of the on-site "exchange" Coulomb matrix elements
into the RPA calculation,
both the collective excitations and the macroscopic dielectric function
of an atomic insulator with the RPA values of the intra-atomic parameters.
If the latter represent an appropriate description of the on-site
correlations,  our method is  already  fully self-consistent.
In the case of the strong on-site correlations,
the short-range contributions to the dielectric response have to be treated
beyond RPA.

Finally, in the limit of large frequencies
($\omega \rightarrow \infty$)
the macroscopic dielectric function (\ref{24}) reduces to the
simple sum of intra-band and inter-band polarizabilities,
with the standard free electron mass expression for the plasma edge
\cite{pines}.

\section { Conclusion}
\mbox{}

In conclusion, we point out the main advantages of the present TB approach
to the dielectric response. Unlike the representation  via the reciprocal
lattice indices, it leads to a simple and physically transparent expression
for the dielectric matrix, which is explicitly resolved in
the particular example of the cubic symmetry.   It also clearly
distinguishes between
the long-range and the local (on-site) contributions to the dielectric
screening,
and gives so the direct insight into the origin of the self-polarization
effects. Furthermore, the present method not only interpolates between
(multiband) conductors and insulators, but also includes systems with
negligible bandwidths as a well-defined asymptotic limit. Just in this
limit we discover an important property of the
dielectric instability in multiband conductors, namely that it proceeds by a
simultaneous softening of one longitudinal and one transverse collective
mode.

 In summary, the approach proposed here facilitates the detailed analysis
of the dielectric response in crystals. It is particularly efficient in the
long wavelength limit, even if a large number of bands has to be retained in
the calculations. In this respect the decisive criterion for the reducibility
of the problem to a finite number of linear equations is the smallness of the
on-site exchange terms (\ref{25}) in comparison with the corresponding
long range contributions. Other on-site terms are not relevant (but they
have to be weak enough in order to justify the starting RPA scheme).
On the other hand, it is clear that our approach has a wide range
of applications to real and model systems in which it is appropriate to
take into account a small number of bands.

\end{sloppypar}

\newpage

\begin{thebibliography}{99}
\bibitem{pines}D. Pines, {\em Elementary Excitations in Solids},
 W.A. Benjamin, Inc. New York, Amsterdam, 1964; D.Pines and  P. Nozieres,
 {\em The Theory of Quantum Liquids}, Vol.1,
 Addison-Wesley Publishing Co., Inc., 1989.
\bibitem{adler}S.L. Adler, Phys. Rev. {\bf 126}, 413 (1962).
\bibitem{wiser}N. Wiser, Phys. Rev. {\bf 129}, 62 (1963).
\bibitem{lcc}S.G. Louie, J.R. Chelikowsky and M.L. Cohen, Phys. Rev. Letters
  {\bf 34}, 155 (1975).
\bibitem{hl}M.S. Hybertsen and S.G. Louie, Phys. Rev. {\bf B35}, 5585 (1987).
\bibitem{onodera}Y. Onodera, Prog. Theor. Physics {\bf 49}, 37 (1973).
\bibitem{sinha}S.K. Sinha, R.P. Gupta and D.L. Price, Phys. Rev. {\bf B9},
 2654 (1974).
\bibitem{barisic}S. Bari\v{s}i\'{c}, Phys. Rev. {\bf B5}, 932  and
  941 (1972).
\bibitem{hanke}W. Hanke, Phys. Rev. {\bf B8}, 4585  and
 4591 (1973).
\bibitem{hanker}For more details see the review article by  W. Hanke,
Adv. Phys. {\bf 27}, 287 (1978).
\bibitem{nozieres}P. Nozieres and D. Pines, Phys. Rev. {\bf 109}, 741
 and 762 (1958) ; Il  Nuovo Cimento {\bf 9}, 470 (1958).
\bibitem{van}J.A..Van Vechten, R.P. Martin Phys. Rev. Lett. {\bf 28},
 446 (1972)
\bibitem{zbb}P.  \v{Z}upanovi\'{c}, A. Bjeli\v{s} and S. Bari\v{s}i\'{c} (to be
published).
\bibitem{cohen}M.H. Cohen and  F. Keller, Phys. Rev. {\bf 99}, 1128 (1955).
\bibitem{tekst1} For simplicity, we assume that the Fermi level
 crosses only one set of degenerate bands ({\em i. e.} not more than
 three bands in a cubic latice). The generalization to an accidental crossing
 of more sets with different symmetries is straightforward.
\bibitem{ll}L. D. Landau and E. M. Lifshitz,
 {\em Quantum Mechanics (Non-Relativistic Theory)}, Pergamon
 Press, 1980.
\bibitem{knox}R.S. Knox, {\em Theory of Excitons},
 Academic Press, New York (1963).
 \bibitem{we}Note that the Wannier excitons, {\em i. e.} the discrete
 bound states at the edges of interband electron-hole
 continua, follow from the inclusion of the extended
 RPA (ladder) terms in the eq.(\ref{3})   [see also  W. Hanke and
 L.J. Sham, Phys. Rev. {\bf B12}, 4501 (1975)]. The straightforward
 generalization of the present approach in this direction shows that the
 dipolar collective modes and the Wannier excitons
 can coexist \cite {zbb}.
\bibitem{anderson}P.W. Anderson, {\em Concepts in Solids},
 W.A. Benjamin, Inc. New York, Amsterdam, 1964.

\end{thebibliography}
\end{document}